\documentclass[twocolumn,showpacs,preprintnumbers,amsmath,amssymb]{revtex4}

\usepackage{graphicx}
\usepackage{dcolumn}
\usepackage{bm}

\begin{document}

\preprint{APS/123-QED}

\title{Deformed special relativity with an invariant minimum speed and its cosmological implications}
\author{Cl\'audio Nassif\\
        (e-mail: {\bf cncruz777@yahoo.com.br})}
 \altaffiliation{{\bf CPFT}: Centro de Pesquisas em F\'isica Te\'orica. Rua Rio de Janeiro 1186/s.1304, 30.160-041, 
Belo Horizonte-MG, Brazil.}

\date{\today}

\begin{abstract}
It will be introduced a new symmetry principle in the space-time
geometry through the elimination of the classical idea of rest, by including a
universal minimum limit of speed in the subatomic world. Such a lowest limit, unattainable by particles,
 represents a preferred reference frame associated with a universal background field that breaks Lorentz
symmetry. Thus the structure of space-time is extended due to the presence
of a vacuum energy density, which leads to a negative pressure at cosmological length
scales. The tiny values of the cosmological constant and the vacuum energy density will be successfully obtained,
 being in good agreement with current observational results.
\end{abstract}

\pacs{98.80.Es; 11.30.Qc}
\maketitle

\section{\label{sec:level1} Introduction}

 Driven by a search for new fundamental symmetries in Nature\cite{1}, we will attempt to implement a uniform background field into
the flat space-time. Such a background field connected to a uniform vacuum energy density represents a preferred reference frame, which
leads us to postulate a universal minimum limit of speed for particles with very large wavelengths (very low energies).

 The hypothesis of a lowest limit of speed in the space-time leads to the following physical reasoning:

  - The plane wave for a free particle is an idealization that is impossible to conceive under physical reality.
   In the event of an idealized plane wave, it would be possible to find with certainty the reference frame
 that cancels its momentum ($p=0$), and the uncertainty on its position would be $\Delta x=\infty$. However,
  the presence of an unattainable minimum limit of speed emerges in order to forbid the ideal case of a plane
 wave ($p=constant$ or $\Delta p=0$). This means that there is no perfect inertial motion ($v=constant$) like a plane wave, except the
 privileged reference frame of a universal background field connected to an unattainable minimum limit of speed $V$, where $p$ would vanish.
  But, since we should consider that such a minimum speed $V$ (universal background frame) is unattainable for the particles with low energies
  (large length scales), actually their momenta can never vanish when one tries to be closer to such a preferred frame ($V$).
  On the other hand, according to Special Relativity (SR), their momenta cannot be infinite since the maximum speed $c$ is also unattainable for such massive
 particles, except the photon ($v=c$) as it is a massless particle. This reasoning allows
 us to think that the electromagnetic radiation (photon:$``c-c''=c$) as well as the massive particle ($``v-v''>V(\neq 0)$ for $v<c$ ) are in
 equal-footing in the sense that it is not possible to find a reference frame at rest ($v_{relative}=0$) for both through any
 speed transformation in a space-time with a maximum and minimum limit of speed. Therefore such a deformed special relativity will
 be denominated {\it Symmetrical Special Relativity} (SSR). We will look for new speed transformations of SSR in the next section.

 The dynamics of particles in the presence of a universal background
reference frame connected to $V$ is within a context of the ideas of
Sciama\cite{2}, Schr\"{o}dinger\cite{3} and Mach\cite{4}, where there should be
an ``absolute" inertial reference frame in relation to which we have the
inertia of all moving bodies. However, we must emphasize that the
approach used here is not classical as machian ideas, since the lowest (unattainable) limit of speed
$V$ plays the role of a privileged (inertial) reference frame of background field instead of the ``inertial" frame of fixed stars.

 It is very curious to notice that the idea of universal background
field was sought in vain by Einstein\cite{5}, motivated firstly by Lorentz. It was Einstein
who coined the term {\it ultra-referential} as the
fundamental aspect of Reality to represent a universal background field\cite{6}. Basing on
such concept, let us call {\it ultra-referential} $S_V$ to be the universal background
field of a fundamental inertial reference frame connected to $V$.

\section{\label{sec:level1}Transformations of space-time and velocity in the presence of the
 ultra-referential $S_V$ }

  SSR should contain 3 postulates, namely:

 1) -{\it the constancy of the speed of light ($c$)}.

 2) -{\it the non-equivalence (asymmetry) of the reference frames in such a space-time,
  i.e., we cannot exchange the speed $v$ (of $S^{\prime }$) for $-v$ (of $S_V$) by the inverse transformations,
  since we cannot find the rest for $S^{\prime}$ ($``v-v''>V$)} (see Fig.1). Such an asymmetry will be clarified later.

 3) -{\it the covariance of the ultra-referential (background frame) $S_V$ connected to
 an unattainable minimum limit of speed $V$} (Fig.1). This third postulate is directly related to the second one above.
  Such a connection will be clarified by studying the new velocity transformations to be obtained soon.

Let us assume the reference frame $S^{\prime}$ with a speed $v$ in relation to the ultra-referential $S_V$ according to Fig. 1.
\begin{figure}
\includegraphics[scale=0.6]{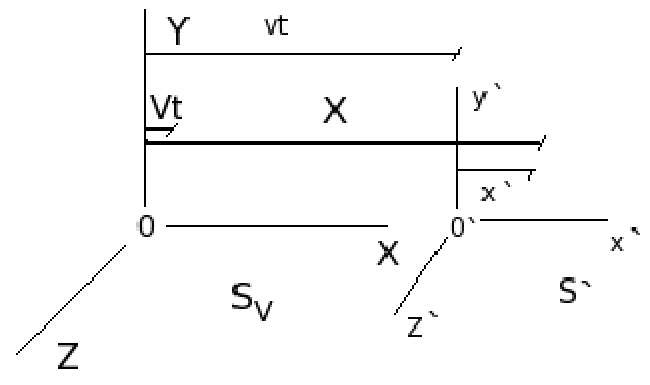}
\caption{$S^{\prime}$ moves with a velocity $v$ with respect to the background field of the covariant ultra-referential $S_V$.
 If $V\rightarrow 0$, $S_V$ is eliminated (empty space) and thus the Galilean frame $S$ takes place, recovering Lorentz transformations.}
\end{figure}

So, to simplify, consider the motion only at one spatial dimension, namely $(1+1)D$ space-time with background field
 $S_V$. So we write the following transformations:

  \begin{equation}
 dx^{\prime}=\Psi(dX-\beta_{*}cdt)=\Psi(dX-vdt+Vdt),
  \end{equation}
where $\beta_{*}=\beta\epsilon=\beta(1-\alpha)$, being $\beta=v/c$ and $\alpha=V/v$, so that
$\beta_{*}\rightarrow 0$ for $v\rightarrow V$ or $\alpha\rightarrow 1$.

 \begin{equation}
 dt^{\prime}=\Psi(dt-\frac{\beta_{*}dX}{c})=\Psi(dt-\frac{vdX}{c^2}+\frac{VdX}{c^2}),
  \end{equation}
being $\vec v=v_x{\bf x}$. We have $\Psi=\frac{\sqrt{1-\alpha^2}}{\sqrt{1-\beta^2}}$. If we make
$V\rightarrow 0$ ($\alpha\rightarrow 0$), we recover Lorentz
transformations, where the ultra-referential $S_V$ is eliminated and simply replaced by the Galilean
frame $S$ at rest for the classical observer.

 In order to get the transformations (1) and (2) above, let us consider the following more general transformations:
$x^{\prime}=\theta\gamma(X-\epsilon_1vt)$ and $t^{\prime}=\theta\gamma(t-\frac{\epsilon_2vX}{c^2})$,
 where $\theta$, $\epsilon_1$ and $\epsilon_2$ are factors (functions) to be determined. We hope all these factors
depend on $\alpha$, such that, for $\alpha\rightarrow 0$ ($V\rightarrow 0$), we recover Lorentz transformations as a particular case
 ($\theta=1$, $\epsilon_1=1$ and $\epsilon_2=1$). By using those transformations to perform
$[c^2t^{\prime 2}-x^{\prime 2}]$, we find the identity: $[c^2t^{\prime 2}-x^{\prime 2}]=
\theta^2\gamma^2[c^2t^2-2\epsilon_1vtX+2\epsilon_2vtX-\epsilon_1^2v^2t^2+\frac{\epsilon_2^2v^2X^2}{c^2}-X^2]$.
 Since the metric tensor is diagonal, the crossed terms must vanish and so we assure that
$\epsilon_1=\epsilon_2=\epsilon$. Due to this fact, the crossed terms ($2\epsilon vtX$) are cancelled between
themselves and finally we obtain $[c^2t^{\prime 2}-x^{\prime 2}]=
 \theta^2\gamma^2(1-\frac{\epsilon^2 v^2}{c^2})[c^2t^2-X^2]$. For $\alpha\rightarrow 0$ ($\epsilon=1$ and
$\theta=1$), we reinstate $[c^2t^{\prime 2}-x^{\prime 2}]=[c^2t^2-x^2]$ of SR. Now we write the following
transformations: $x^{\prime}=\theta\gamma(X-\epsilon vt)\equiv\theta\gamma(X-vt+\delta)$ and
$t^{\prime}=\theta\gamma(t-\frac{\epsilon vX}{c^2})\equiv\theta\gamma(t-\frac{vX}{c^2}+\Delta)$, where
we assume $\delta=\delta(V)$ and $\Delta=\Delta(V)$, so that $\delta =\Delta=0$ for $V\rightarrow 0$, which implies $\epsilon=1$.
 So from such transformations we extract: $-vt+\delta(V)\equiv-\epsilon vt$ and
$-\frac{vX}{c^2}+\Delta(V)\equiv-\frac{\epsilon vX}{c^2}$, from where we obtain
 $\epsilon=(1-\frac{\delta(V)}{vt})=(1-\frac{c^2\Delta(V)}{vX})$. As $\epsilon$ is a dimensionaless factor,
we immediately conclude that $\delta(V)=Vt$ and $\Delta(V)=\frac{VX}{c^2}$, so that we find
$\epsilon=(1-\frac{V}{v})=(1-\alpha)$. On the other hand, we can determine $\theta$ as follows: $\theta$ is a
function of $\alpha$ ($\theta(\alpha)$), such that $\theta=1$ for $\alpha=0$, which also leads to $\epsilon=1$ in
order to recover Lorentz transformations. So, as $\epsilon$ depends on $\alpha$, we conclude that $\theta$ can
also be expressed in terms of $\epsilon$, namely $\theta=\theta(\epsilon)=\theta[(1-\alpha)]$, where
$\epsilon=(1-\alpha)$. Therefore we can write $\theta=\theta[(1-\alpha)]=[f(\alpha)(1-\alpha)]^k$, where the
exponent $k>0$. The function $f(\alpha)$ and $k$ will be estimated by satisfying the following conditions:

i) as $\theta=1$ for $\alpha=0$ ($V=0$), this implies $f(0)=1$.

ii) the function $\theta\gamma =
\frac{[f(\alpha)(1-\alpha)]^k}{(1-\beta^2)^{\frac{1}{2}}}=\frac{[f(\alpha)(1-\alpha)]^k}
{[(1+\beta)(1-\beta)]^{\frac{1}{2}}}$ should have a symmetrical behavior, that is to say it goes to zero closer
to $V$ ($\alpha\rightarrow 1$) in the same way it goes to infinite closer to $c$ ($\beta\rightarrow 1$). In other words, this
means that the numerator of the function $\theta\gamma$, which depends on $\alpha$ should have the same shape of its denumerator,
  which depends on $\beta$. Due to such conditions, we naturally conclude that $k=1/2$ and
$f(\alpha)=(1+\alpha)$, so that $\theta\gamma=
\frac{[(1+\alpha)(1-\alpha)]^{\frac{1}{2}}}{[(1+\beta)(1-\beta)]^{\frac{1}{2}}}=
\frac{(1-\alpha^2)^{\frac{1}{2}}}{(1-\beta^2)^\frac{1}{2}}=\frac{\sqrt{1-V^2/v^2}}{\sqrt{1-v^2/c^2}}=\Psi$, where
$\theta=\sqrt{1-\alpha^2}=\sqrt{1-V^2/v^2}$.

The transformations shown in (1) and (2) are the direct transformations
from $S_V$ [$X^{\mu}=(X,ict)$] to $S^{\prime}$ [$x^{\prime\nu}=(x^{\prime},ict^{\prime})$], where
we have $x^{\prime\nu}=\Omega^{\nu}_{\mu} X^{\mu}$ ($x^{\prime}=\Omega X$),
so that we obtain the following matrix of transformation:

\begin{equation}
\displaystyle\Omega=
\begin{pmatrix}
\Psi & i\beta (1-\alpha)\Psi \\
-i\beta (1-\alpha)\Psi & \Psi
\end{pmatrix},
\end{equation}
such that $\Omega\rightarrow\ L$ (Lorentz matrix of rotation) for $\alpha\rightarrow 0$
($\Psi\rightarrow\gamma$).

We obtain $det\Omega =\frac{(1-\alpha^2)}{(1-\beta^2)}[1-\beta^2(1-\alpha)^2]$, where $0<det\Omega<1$. Since
$V$ ($S_V$) is unattainable ($v>V)$, this assures that $\alpha=V/v<1$ and therefore the matrix $\Omega$
admits inverse ($det\Omega\neq 0$ $(>0)$). However $\Omega$ is a non-orthogonal matrix
($det\Omega\neq\pm 1$) and so it does not represent a rotation matrix ($det\Omega\neq 1$) in such a space-time
due to the presence of the privileged frame of background field $S_V$ that breaks strongly the invariance of the norm of the
4-vector (limit $v\rightarrow V$ in (15) or (16)).  Actually such an effect ($det\Omega\approx 0$ for $\alpha\approx 1$) emerges from
a new relativistic physics of SSR for treating much lower energies at infrared regime closer to $S_V$ (very large wavelengths).

 We notice that $det\Omega$ is a function of the speed $v$ with respect to $S_V$. In the approximation for
$v>>V$ ($\alpha\approx 0$), we obtain $det\Omega\approx 1$ and so we practically reinstate the rotation behavior
of Lorentz matrix as a particular regime for higher energies. If we make $V\rightarrow 0$ ($\alpha\rightarrow 0$),
 we recover $det\Omega=1$.

The inverse transformations (from $S^{\prime}$ to $S_V$) are

 \begin{equation}
 dX=\Psi^{\prime}(dx^{\prime}+\beta_{*}cdt^{\prime})=\Psi^{\prime}(dx^{\prime}+vdt^{\prime}-Vdt^{\prime}),
  \end{equation}

 \begin{equation}
 dt=\Psi^{\prime}(dt^{\prime}+\frac{\beta_{*}
 dx^{\prime}}{c})=\Psi^{\prime}(dt^{\prime}+\frac{vdx^{\prime}}{c^2}-
\frac{Vdx^{\prime}}{c^2}).
  \end{equation}

In matrix form, we have the inverse transformation $X^{\mu}=\Omega^{\mu}_{\nu} x^{\prime\nu}$
 ($X=\Omega^{-1}x^{\prime}$), so that the inverse matrix is

\begin{equation}
\displaystyle\Omega^{-1}=
\begin{pmatrix}
\Psi^{\prime} & -i\beta (1-\alpha)\Psi^{\prime} \\
 i\beta (1-\alpha)\Psi^{\prime} & \Psi^{\prime}
\end{pmatrix},
\end{equation}
where we can show that $\Psi^{\prime}$=$\Psi^{-1}/[1-\beta^2(1-\alpha)^2]$, so that we must satisfy $\Omega^{-1}\Omega=I$.

 Indeed we have $\Psi^{\prime}\neq\Psi$ and therefore $\Omega^{-1}\neq\Omega^T$. This non-orthogonal aspect of
$\Omega$ has
an important physical implication. In order to understand such an implication, let us consider firstly the
orthogonal (e.g: rotation) aspect of Lorentz matrix in SR. Under SR, we
have $\alpha=0$, so that $\Psi^{\prime}\rightarrow\gamma^{\prime}=\gamma=(1-\beta^2)^{-1/2}$.
 This symmetry ($\gamma^{\prime}=\gamma$, $L^{-1}=L^T$) happens because the Galilean reference
frames allow us to exchange the speed $v$ (of $S^{\prime}$) for $-v$ (of $S$) when we are at rest at
$S^{\prime}$. However, under SSR, since there is no rest at $S^{\prime}$, we cannot exchange $v$ (of $S^{\prime}$) for $-v$ (of $S_V$)
due to that asymmetry ($\Psi^{\prime}\neq\Psi$, $\Omega^{-1}\neq\Omega^T$). Due to this fact,
$S_V$ must be covariant, namely $V$ remains invariant for any change of reference frame in such a space-time. Thus we
can notice that the paradox of twins, which appears due to that symmetry by
exchange of $v$ for $-v$ in SR should be naturally eliminated in SSR where only the
reference frame $S^{\prime}$ can move with respect to $S_V$. So $S_V$ remains
covariant (invariant for any change of reference frame). Such a covariance will be verified soon.

  We have $det\Omega=\Psi^2[1-\beta^2(1-\alpha)^2]\Rightarrow [(det\Omega)\Psi^{-2}]=[1-\beta^2(1-\alpha)^2]$. So
we can alternatively write $\Psi^{\prime}$=$\Psi^{-1}/[1-\beta^2(1-\alpha)^2]=\Psi^{-1}/[(det\Omega)\Psi^{-2}]
=\Psi/det\Omega$. By inserting this result in (6) to replace $\Psi^{\prime}$, we obtain the relationship
between the inverse matrix and the transposed matrix of $\Omega$, namely $\Omega^{-1}=\Omega^T/det\Omega$. Indeed
$\Omega$ is a non-orthogonal matrix, since we have $det\Omega\neq\pm 1$.

 By dividing (1) by (2), we obtain the following speed transformation:

   \begin{equation}
  v_{Rel}=\frac{v^{\prime}-v+V}
{1-\frac{v^{\prime}v}{c^2}+\frac{v^{\prime}V}{c^2}},
   \end{equation}
 where we have considered $v_{Rel}=v_{Relative}\equiv dx^{\prime}/dt^{\prime}$
 and $v^{\prime}\equiv dX/dt$.  $v^{\prime}$ and $v$ are given with
 respect to $S_V$, and $v_{Rel}$ is related between them. Let us consider
 $v^{\prime}>v$. (see Fig.2)

\begin{figure}
\includegraphics[scale=0.6]{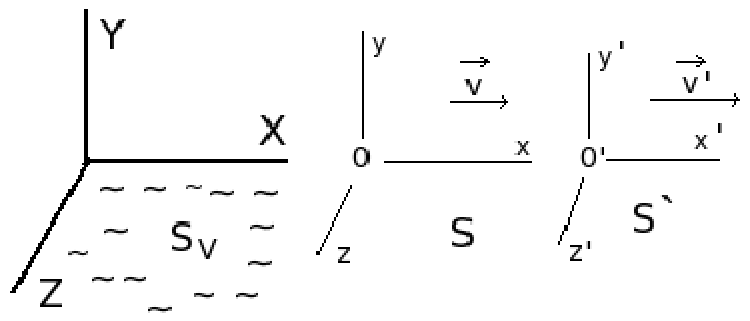}
\caption{$S_V$ is the covariant ultra-referential of background field. $S$ represents the
reference frame for a massive particle with speed $v$ in relation to $S_V$, where $V<v<c$.
  $S^{\prime}$ represents the reference frame for a massive particle with speed $v^{\prime}$
in relation to $S_V$. In this case, we consider $V<v\leq v^{\prime}\leq c$.\/}
\end{figure}

 If $V\rightarrow 0$, the transformation (7) recovers the Lorentz
 velocity transformation where $v^{\prime}$ and $v$ are given in relation to a
 certain Galilean frame $S_0$ at rest. Since (7) implements the ultra-referential
 $S_V$, the speeds $v^{\prime}$ and $v$ are now given with respect to $S_V$,
  which is covariant (absolute). Such a covariance is verified if we assume that
 $v^{\prime}=v=V$ in (7). Thus, for this case, we obtain $v_{Rel}=``V-V''=V$.

  Let us also consider the following cases in (7):

 {\bf a)} $v^{\prime}=c$ and $v\leq c\Rightarrow v_{Rel}=c$. This
 just verifies the well-known invariance of $c$.

 {\bf b)} if $v^{\prime}>v(=V)\Rightarrow v_{Rel}=``v^{\prime}-V"=v^{\prime}$. For
 example, if $v^{\prime}=2V$ and $v=V$ $\Rightarrow v_{Rel}=``2V-V"=2V$. This
 means that $V$ really has no influence on the speed of the particles. So $V$ works as if
 it were an ``{\it absolute zero of movement}'', being invariant and having the same value
in all directions of space of the isotropic background field.

 {\bf c)} if $v^{\prime}=v$ $\Rightarrow v_{Rel}=``v-v''$($\neq 0)$
$=\frac{V}{1-\frac{v^2}{c^2}(1-\frac{V}{v})}$. From ({\bf c}) let us
consider two specific cases, namely:

  -$c_1$) assuming $v=V\Rightarrow v_{Rel}=``V-V"=V$ as verified before.

  -$c_2$) if $v=c\Rightarrow v_{Rel}=c$,
 where we have the interval $V\leq v_{Rel}\leq c$ for $V\leq v\leq c$.

This last case ({\bf c}) shows us in fact that it is impossible to find the
rest for the particle on its own reference frame $S^{\prime}$, where
$v_{Rel}(v)$ ($\equiv\Delta v(v)$) is a function that increases with the increasing of $v$ . However,
 if we make $V\rightarrow 0$, then we would have $v_{Rel}\equiv\Delta v=0$ and therefore
it would be possible to find the rest for $S^{\prime}$, which would become simply a Galilean
reference frame of SR.

 By dividing (4) by (5), we obtain

 \begin{equation}
  v_{Rel}=\frac{v^{\prime}+v-V}
{1+\frac{v^{\prime}v}{c^2}-\frac{v^{\prime}V}{c^2}}
   \end{equation}

 In (8), if $v^{\prime}=v=V\Rightarrow ``V+V''=V$. Indeed $V$ is
 invariant, working like an {\it absolute zero state} in SSR. If
 $v^{\prime}=c$ and $v\leq c$, this implies $v_{Rel}=c$. For $v^{\prime}>V$
 and considering $v=V$, this leads to $v_{Rel}=v^{\prime}$. As a specific example, if $v^{\prime}=2V$
 and assuming $v=V$, we would have $v_{Rel} =``2V+V''=2V$. And if
 $v^{\prime}=v\Rightarrow v_{Rel}=``v+v"=\frac{2v-V}{1+\frac{v^2}{c^2}(1-\frac{V}{v})}$.
  In newtonian regime ($V<<v<<c$), we recover $v_{Rel}=``v+v"=2v$. In relativistic (einsteinian)
regime ($v\rightarrow c$), we reinstate Lorentz transformation for this case ($v^{\prime}=v$),
 i.e., $v_{Rel}=``v+v"=2v/(1+v^2/c^2)$.

 By joining both transformations (7) and (8) into just one, we write the following compact form:

\begin{equation}
  v_{Rel}=\frac{v^{\prime}\mp\epsilon v}
{1\mp\frac{v^{\prime}\epsilon v}{c^2}}=\frac{v^{\prime}\mp v(1-\alpha)}
{1\mp\frac{v^{\prime}v(1-\alpha)}{c^2}}=\frac{v^{\prime}\mp v\pm V}
{1\mp \frac{v^{\prime}v}{c^2}\pm \frac{v^{\prime}V}{c^2}},
\end{equation}
being $\alpha=V/v$ and $\epsilon=(1-\alpha)$. For $\alpha=0$ ($V=0$) or $\epsilon=1$, we recover Lorentz speed
transformations.

  Transformations for $(3+1)D$ and also a new group algebra for SSR will be treated elsewhere.

\section{Covariance of the Maxwell wave equation in the presence of the ultra-referential $S_V$}

Let us assume a light ray emitted from the frame $S^{\prime}$. Its equation of electrical wave
at this reference frame is

\begin{equation}
\frac{\partial^2\vec E(x^{\prime},t^{\prime})}{\partial x^{\prime 2}}-
\frac{1}{c^2}\frac{\partial^2\vec E(x^{\prime},t^{\prime})}
{\partial t^{\prime 2}}=0
\end{equation}

 As it is already known, when we make the exchange by conjugation on the
spatial and temporal coordinates, we obtain respectively the following
operators: $X\rightarrow\partial/\partial t$ and $t\rightarrow\partial/\partial X$;
also $x^{\prime}\rightarrow\partial/\partial t^{\prime}$ and $t^{\prime}\rightarrow\partial/
\partial x^{\prime}$. Thus the transformations (1) and (2) for such differential operators are

\begin{equation}
\frac{\partial}{\partial t^{\prime}}=
\Psi[\frac{\partial}{\partial t}-\beta
 c(1-\alpha)\frac{\partial}{\partial X})],
\end{equation}

\begin{equation}
\frac{\partial}{\partial x^{\prime}}=
\Psi[\frac{\partial}{\partial X}-\frac{\beta}{c}(1-\alpha)\frac{\partial}{\partial t})],
\end{equation}
where $\beta =v/c$ and $\alpha=V/v$ (see Fig.1).

By squaring (11) and (12), inserting into (10) and after performing the calculations,
we will finally obtain

\begin{equation}
det\Omega\left(\frac{\partial^2\vec E}{\partial X^2}
-\frac{1}{c^2}\frac{\partial^2\vec E}{\partial t^2}\right)=0,
\end{equation}
where $det\Omega =\Psi^2[1-\beta^2(1-\alpha)^2]$ (see (3)).

  As the ultra-referential $S_V$ is definitely inaccessible for any particle, we always have
$\alpha<1$ (or $v>V$), which always implies $det\Omega=\Psi^2[1-\beta^2(1-\alpha)^2]>0$. And as we
already have shown in the last section, such a result is in agreement with the fact that we must
have $det\Omega>0$. Therefore this will always assure

  \begin{equation}
 \frac{\partial^2\vec E}{\partial X^2}
-\frac{1}{c^2}\frac{\partial^2\vec E}{\partial t^2}=0
 \end{equation}

By comparing (14) with (10), we verify the covariance of the electromagnetic wave equation propagating in the background field of
the ultra-referential $S_V$.

\section{\label{sec:level1} The flat space-time and the ultra-referential $S_V$}

Let us consider the ultra-referential $S_V$ as a uniform background field that fills
the whole flat space-time as a perfect fluid, playing the role of a kind of de-Sitter
 (dS) space-time\cite{7} shown in the next section ($\Lambda>0)$. So let us define the following metric:
\begin{equation}
ds^2=\Theta g_{\mu\nu}dx^{\mu}dx^{\nu},
\end{equation}
where $g_{\mu\nu}$ is the well-known Minkowski metric. $\Theta$ is a
scale factor that increases for very large wavelengths (cosmological scales) governed by vacuum (dS), that is to say for much
lower energies, where we have $\Theta\rightarrow\infty$. On the other hand, $\Theta$ decreases to 1 for smaller scales of length, namely for higher
energies ($\Theta\rightarrow 1$) where dS space-time approximates to the Minkowski metric as a special case.  $\Theta$ breaks strongly the invariance of $ds$
only for very large distances governed by vacuum of the ultra-referential $S_V$. For smaller scales of
length governed by matter, we naturally restore Lorentz symmetry and the invariance of $ds$. Following such considerations,
 let us consider $\Theta$ to be a function of speed $v$ with respect to the background field-$S_V$, namely:
\begin{equation}
\Theta =\Theta (v)=\frac{1}{(1-\frac{V^2}{v^2})},
\end{equation}
such that $\Theta\approx 1$ for $v>>V$ (Lorentz symmetry regime) and
$\Theta\rightarrow\infty$ for $v\rightarrow V$ (regime of ultra-referential $S_V$ that
breaks strongly $ds$ invariance, so that $ds\rightarrow\infty$).

 The total energy $E$ of a particle in $S_V$ is
\begin{equation}
E=\theta(\gamma mc^2)=\Psi mc^2 =mc^2\frac{\sqrt{1-\frac{V^2}{v^2}}}{\sqrt{1-\frac{v^2}{c^2}}},
\end{equation}
where $\theta =\Theta^{-1/2}=\sqrt{1-\alpha^2}$ and $\gamma =1/\sqrt{1-\beta^2}$, being $\alpha=V/v$
and $\beta=v/c$. $v$ is given in relation to $S_V$.

 In (17), we observe that $E\rightarrow 0$ for $v\rightarrow V$ ($S_V$). For the case
$v=v_0=\sqrt{cV}$, we obtain $\theta\gamma =\Psi(v_0) =1\Rightarrow E=mc^2$.
  Actually, as a massive particle always has motion $v$ ($V(S_V)<v<c$) with respect to the unattainable
ultra-referential $S_V$, its proper energy $mc^2$ requires a non-zero motion $v(=v_0)$ in relation to $S_V$ (see Fig.3).

\begin{figure}
\begin{center}
\includegraphics[scale=0.6]{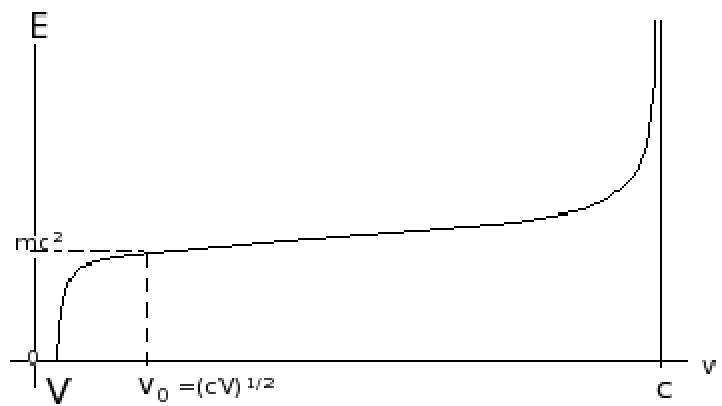}
\end{center}
\caption{\it $v_0$ represents the speed in relation to $S_V$, from where we get the proper energy of the particle
 ($E_0=mc^2$), being $\Psi_0=\Psi(v_0)=1$. For $v<<v_0$ or closer to $S_V$ ($v\rightarrow V$), a new relativistic correction on energy arises,
 so that $E\rightarrow 0$.}
\end{figure}

 The momentum of the particle in relation to $S_V$ is
\begin{equation}
\vec P = m\vec v\frac{\sqrt{1-\frac{V^2}{v^2}}}{\sqrt{1-\frac{v^2}{c^2}}}.
\end{equation}

 From (17) and (18), we show the following energy-momentum relation: $c^2\vec P^2=E^2-m^2c^4(1-\frac{V^2}{v^2})$.

 The de-Broglie wavelength of the particle in $S_V$ is due to its motion $v$ with respect to $S_V$, namely:

\begin{equation}
 \lambda=\frac{h}{P}=\frac{h}{mv}\frac{\sqrt{1-\frac{v^2}{c^2}}}{\sqrt{1-\frac{V^2}{v^2}}},
 \end{equation}
from where we have used the momentum (18) given with respect to $S_V$.

If $v\rightarrow c\Rightarrow\lambda\rightarrow 0$ ({\it spatial
contraction}), and if $v\rightarrow V~(S_V)\Rightarrow\lambda\rightarrow\infty$
({\it spatial dilation to the infinite}, breaking strongly Lorentz symmetry in SSR), which means we have cosmological
wavelengths. This leads to $\Theta\rightarrow\infty$ (see (16)).

\section{\label{sec:level1} Cosmological implications}

\subsection{Energy-momentum tensor in the presence of the ultra-referential-$S_V$}

  Let us write the 4-velocity in the presence of $S_V$, as follows:
   \begin{equation}
 U^{\mu}=\left[\frac{\sqrt{1-\frac{V^2}{v^2}}}{\sqrt{1-\frac{v^2}{c^2}}}~ , ~
\frac{v_{\alpha}\sqrt{1-\frac{V^2}{v^2}}}{c\sqrt{1-\frac{v^2}{c^2}}}\right],
   \end{equation}
where $\mu=0,1,2,3$ and $\alpha=1,2,3$. If $V\rightarrow 0$, we recover the 4-velocity of SR.

The well-known energy-momentum tensor to deal with perfect fluid is of the form
   \begin{equation}
  T^{\mu\nu}=(p+\epsilon)U^{\mu}U^{\nu} - pg^{\mu\nu},
   \end{equation}
where $U^{\mu}$ is given in (20). $p$ represents a pressure and $\epsilon$ an energy density.

   From (20) and (21), performing the new component $T^{00}$, we obtain
   \begin{equation}
  T^{00}=\frac{\epsilon(1-\frac{V^2}{v^2})+p(\frac{v^2}{c^2}-\frac{V^2}{v^2})}{(1-\frac{v^2}{c^2})}
   \end{equation}

 If $V\rightarrow 0$, we recover the old component $T^{00}$.

Now, as we are interested only in obtaining $T^{00}$ in absence of matter, i.e.,
 the vacuum limit connected to the ultra-referential $S_V$, we perform the limit of (22) as follows:
 \begin{equation}
 lim_{v\rightarrow V} T^{00}= T^{00}_{vacuum}=\frac{p(\frac{V^2}{c^2}-1)}{(1-\frac{V^2}{c^2})}= -p.
  \end{equation}

 From (22), we notice that the term $\epsilon\gamma^2(1-V^2/v^2)$ for representing matter vanishes naturally
 in the limit of vacuum-$S_V$ ($v\rightarrow V$), and therefore just the contribution of vacuum prevails. As we always
 must have $T^{00}>0$, we get $p<0$ in (23). This implies a negative pressure for vacuum energy density of the
ultra-referential $S_V$. So we verify that a negative pressure emerges naturally from such a new tensor in the limit of $S_V$.

  We can obtain $T^{\mu\nu}_{vacuum}$ by calculating the limit of vacuum-$S_V$ for (21), by considering (20), as follows:
 \begin{equation}
 T^{\mu\nu}_{vacuum}= lim_{v\rightarrow V}T^{\mu\nu}= -pg^{\mu\nu},
 \end{equation}
 where we conclude that $\epsilon=-p$. In (20), we see that the new 4-velocity vanishes in the limit of the vacuum-$S_V$ ($v\rightarrow V$),
 namely $U^{\mu}_{vac.}=(0,0)$. So $T^{\mu\nu}_{vac.}$ is in fact a diagonalized tensor as we hope to be. The vacuum-$S_V$ inherent to
such a space-time works like a {\it sui generis} fluid in equilibrium with negative pressure, leading to a cosmological anti-gravity.

\subsection{The cosmological constant $\Lambda$}

 The well-known relation\cite{16} between the cosmological constant $\Lambda$ and the vacuum energy density $\rho_{(\Lambda)}$ is
\begin{equation}
\rho_{(\Lambda)}=\frac{\Lambda c^2}{8\pi G}
\end{equation}

Let us consider a spherical universe with its Hubble radius filled by a uniform vacuum energy density. On the surface of such a
sphere (frontier of the observable universe), the bodies (galaxies) experiment an accelerated expansion (anti-gravity) due to
the whole ``dark mass" of vacuum inside the sphere. So we could think that each galaxy is a proof body interacting with that big sphere
like in the simple case of two bodies interaction, however we need to show that there is an anti-gravitational interaction. But before
this, let us first start from the well-known simple model of a massive particle that escapes from a classical gravitational potential $\phi$
on the surface of a spherical mass, namely $E=mc^2(1-v^2/c^2)^{-1/2}\equiv mc^2(1+\phi/c^2)$, where $E$ is its relativistic energy.
 Here the interval of velocity $0\leq v<c$ is associated with the interval of potential $0\leq\phi<\infty$, where we stipulate $\phi>0$ to be
the attractive potential. Now it is important to notice that the strong influence of the background field (vacuum energy) connected to the ultra-referential
$S_V$ leads to a strong repulsive (negative) gravitational potential ($\phi<<0$) for very low energies ($E\rightarrow 0$). This modified (non-classical)
aspect of gravitation\cite{8} prevails only for cosmological scales of length governed by vacuum-$S_V$. In order to see such an aspect, we write the
approximation for much lower energies in (17), as follows:
\begin{equation}
E\approx mc^2(1-V^2/v^2)^{1/2}\equiv mc^2(1+\phi/c^2),
\end{equation}
where, for $E\rightarrow 0$, this implies $v\rightarrow V$, which leads to $\phi\rightarrow -c^2$. So, the non-classical minimum potential $\phi(=-c^2)$
connected to vacuum $S_V$ is responsible for the cosmological anti-gravity. We interpret this result assuming that only an exotic ``particle" of the
vacuum-$S_V$ can escape from the anti-gravity ($\phi=-c^2$) generated by the own cosmological vacuum-$S_V$. Therefore, ordinary proof bodies like galaxies
and any matter on the surface of the sphere cannot escape from its anti-gravity, being accelerated for away.

 According to (26), we should note that such an exotic ``particle" of vacuum (at $S_V$) has an infinite mass since we should consider $v=V$ ($\theta=0$)
in order to have a finite value of $E$, other than the photon ($v=c$) with a null mass (see (17)). So
we conclude that an exotic ``particle" of vacuum works like a counterparty of the photon, namely an infinitely massive boson.

 In (26) the most negative potential (for $v=V$) related to the cosmological constant (vacuum energy) is

\begin{equation}
\phi_{\Lambda}=\phi(V)=-c^2
\end{equation}

  Such a negative potential depends directly on $\Lambda$, namely,
  $\phi_{\Lambda}=\phi(\Lambda)=\phi(V)=-c^2$. To show that, let us
consider that simple model of spherical universe with a radius $R_u$, being filled by
a uniform vacuum energy density $\rho_{(\Lambda)}$, so that the total vacuum energy inside the sphere is
$E_{\Lambda}=\rho_{(\Lambda)}V_u=-pV_u=M_{\Lambda}c^2$. $V_u$ is its volume and $M_{\Lambda}$ is the total
dark mass associated with the dark energy for $\Lambda$ ($w=-1$). Therefore the repulsive gravitational
potential on the surface of such a sphere is

\begin{equation}
\phi_{\Lambda}=-\frac{GM_{\Lambda}}{R_u}=-\frac{G\rho_{(\Lambda)}V_u}{R_uc^2}=\frac{4\pi GpR_u^2}{3c^2},
\end{equation}
where $p=-\rho_{(\Lambda)}$, with $w=-1$.

 By introducing (25) into (28), we find
\begin{equation}
\phi_{\Lambda}=\phi(\Lambda)=-\frac{\Lambda R_u^2}{6}
\end{equation}

Finally, by comparing (29) with (27), we extract

\begin{equation}
\Lambda=\frac{6c^2}{R_u^2},
\end{equation}
where $\Lambda S_u=24\pi c^2$, being $S_u=4\pi R_u^2$.

And also by comparing (28) with (27), we have

\begin{equation}
\rho_{(\Lambda)}=-p=\frac{3c^4}{4\pi G R_u^2},
\end{equation}
where $\rho_{(\Lambda)} S_u=3c^4/G$. (31) and (30) satisfy (25).

 In (30), $\Lambda$ is a kind of {\it cosmological scalar field}, extending the old concept of
Einstein about the cosmological constant for stationary universe. From (30), by considering the Hubble radius, with
$R_{u}=R_{H_0}\sim 10^{26}m$, we obtain $\Lambda=\Lambda_0\sim (10^{17}m^2s^{-2}/10^{52}m^2)\sim 10^{-35}s^{-2}$.
To be more accurate, we know the age of the universe $T_0=13.7$ Gyr, being $R_{H_0}=cT_0\approx
1.3\times 10^{26}m$, which leads to $\Lambda_0\approx 3\times 10^{-35}s^{-2}$. This tiny positive value is very close to
the observational results\cite{9}\cite{10}\cite{11}\cite{12}\cite{13}. The vacuum energy
density\cite{14}\cite{15} given in (31) for $R_{H_0}$ is $\rho_{(\Lambda_0)}\approx
2\times 10^{-29}g/cm^{3}$, which is also in agreement with observations. For scale of the Planck length, where
$R_{u}=l_P=(G\hbar/c^3)^{1/2}$, from (30) we find $\Lambda=\Lambda_P=6c^5/G\hbar\sim 10^{87}s^{-2}$, and from (31)
$\rho_{(\Lambda)}=\rho_{(\Lambda_P)}=T^{00}_{vac.P}=\Lambda_P c^2/8\pi G=3c^7/4\pi G^2\hbar\sim 10^{113}J/m^3
(=3c^4/4\pi l_P^2G\sim 10^{43}kgf/S_P\sim 10^{108}atm\sim 10^{93}g/cm^3)$. So just at that past time, $\Lambda_P$ or
$\rho_{(\Lambda_P)}$ played the role of an inflationary vacuum field with 122 orders of magnitude\cite{16} beyond of
those ones ($\Lambda_0$ and $\rho_{(\Lambda_0)}$) for the present time.

It must be stressed that our assumption for obtaining the tiny positive value of $\Lambda$ starts from first principles related to
a new symmetry in the spacetime, i.e., the idea of a background reference frame for the vacuum energy connected to an invariant 
and unattainable minimum speed given in the quantum world.

\end{document}